\documentclass[aps,pra,superscriptaddress,twocolumn]{revtex4-1}

\usepackage{graphicx}
\usepackage{color}
\usepackage{epstopdf}
\usepackage{amsmath}

\begin{document}

\title{Non-classical higher-order photon correlations with a quantum dot strongly coupled to a photonic-crystal nanocavity}

\author{Armand Rundquist}
\email{armandhr@stanford.edu}
\affiliation{E.\ L.\ Ginzton Laboratory, Stanford University, Stanford, CA}

\author{Michal Bajcsy}
\email{mbajcsy@uwaterloo.ca}
\affiliation{E.\ L.\ Ginzton Laboratory, Stanford University, Stanford, CA}
\affiliation{Institute for Quantum Computing, University of Waterloo, Waterloo, ON}

\author{Arka Majumdar}
\affiliation{E.\ L.\ Ginzton Laboratory, Stanford University, Stanford, CA}
\affiliation{Department of Physics, University of California, Berkeley, CA}
\affiliation{Department of Electrical Engineering, University of Washington, Seattle, WA}

\author{Tomas Sarmiento}
\affiliation{E.\ L.\ Ginzton Laboratory, Stanford University, Stanford, CA}

\author{Kevin Fischer}
\affiliation{E.\ L.\ Ginzton Laboratory, Stanford University, Stanford, CA}

\author{Konstantinos G.\ Lagoudakis}
\affiliation{E.\ L.\ Ginzton Laboratory, Stanford University, Stanford, CA}

\author{Sonia Buckley}
\affiliation{E.\ L.\ Ginzton Laboratory, Stanford University, Stanford, CA}

\author{Alexander Y. Piggott}
\affiliation{E.\ L.\ Ginzton Laboratory, Stanford University, Stanford, CA}

\author{Jelena Vu\v{c}kovi\'{c}}
\affiliation{E.\ L.\ Ginzton Laboratory, Stanford University, Stanford, CA}
\date{\today}

\begin{abstract}
We use the third- and fourth-order autocorrelation functions $g^{(3)}(\tau_1,\tau_2)$ and $g^{(4)}(\tau_1,\tau_2, \tau_3)$ to detect the non-classical character of the light transmitted through a photonic-crystal nanocavity containing a strongly-coupled quantum dot probed with a train of coherent light pulses. We contrast the value of $g^{(3)}(0, 0)$ with the conventionally used $g^{(2)}(0)$ and demonstrate that in addition to being necessary for detecting two-photon states emitted by a low-intensity source, $g^{(3)}$ provides a more clear indication of the non-classical character of a light source. We also present preliminary data that demonstrates bunching in the fourth-order autocorrelation function $g^{(4)}(\tau_1,\tau_2, \tau_3)$ as the first step toward detecting three-photon states.
\end{abstract}

\maketitle

A strongly-coupled quantum dot--cavity system can produce non-classical light  by filtering the input stream of photons coming from a classical coherent light source through mechanisms described as `photon blockade' \cite{Kimble2005, Faraon2008a} and `photon-induced tunneling' \cite{Faraon2008a, Reinhard2012}. Recent proposals \cite{Majumdar2012a, Munoz2013} have extended the concept of photon blockade from single photons to two-photon Fock state generation by coupling the probe laser to the second manifold of the Jaynes-Cummings ladder via a two-photon transition \cite{Kubanek2008}. This approach can potentially be further generalized to create third- and higher-order photon states inside the cavity through multi-photon transitions to the corresponding manifold. Following our proposal \cite{Majumdar2012a}, we report the probing of these multi-photon transitions into the higher manifolds of the Jaynes-Cummings ladder of a strongly coupled quantum dot--photonic crystal nanocavity system \cite{Faraon2008a} by measuring the third-order autocorrelation function ($g^{(3)}(\tau_1,\tau_2)$) of a probe laser transmitted through such a system. Prior to this work, higher-order photon correlations had been measured for thermal \cite{Assman2009, Stevens2010, Zhou2010, Ma2011} and laser \cite{Wiersig2009} sources, relying on the strong excitation and high count rates available in these systems. Very recently $g^{(3)}$ measurements of the fluorescence from a single quantum dot weakly coupled to a microcavity were reported as well \cite{Stevens2014}. However, in the low-intensity, strongly-coupled regime of cavity quantum electrodynamics, such correlations have only been measured in an atomic system \cite{Koch2011}. Therefore, this work constitutes a significant step towards implementing a solid-state non-classical light source of photon number states.

One of the benchmarks used to characterize a source of single photons is the measurement of the the second-order autocorrelation function $g^{(2)}(\tau)={\langle a^{\dag} a^{\dag}(\tau)a(\tau) a\rangle \over \langle  a^{\dag}a\rangle^2}$ \cite{Glauber1963} at $\tau=0$, which quantifies the suppression of multi-photon states. In actual experiments, the value of $g^{(2)}(0)$ for a light source is usually estimated from a Hanbury-Brown and Twiss (HBT) setup that measures coincidence counts between two single photon counting modules (SPCMs). A classical coherent light source will produce photons with Poisson statistics ($g^{(2)}(0)=1$), while a source whose output contains at most one photon at a time will produce $g^{(2)}(0)=0$. More generally, photons from a ``sub-Poissonian" light source -- in which the single photon component dominates over the multi-photon states -- will be anti-bunched and yield $g^{(2)}(0)<1$.

In theory, the second-order autocorrelation function can also be used to identify a two-photon source, since a pure two-photon Fock state will have $g^{(2)}(0)={1/2}$. However, most experimental demonstrations of non-classical light sources result in low-intensity (sparse) output, i.e.\ the source is outputting zero photons most of the time. While this does not affect the value of $g^{(2)}(0)$ for a single-photon source, a perfect but low-intensity two-photon source outputting the state $\psi \approx \sqrt{1-\varepsilon^2}\vert 0 \rangle + \varepsilon \vert 2 \rangle$, with $\varepsilon \ll 1$, will give $g^{(2)}(0)\approx{1/2 \varepsilon^2}$ (see Appendix A). A similar argument can be made for any perfect but sparse $n$-photon source, which illustrates the difficulty of quantitatively distinguishing between various multi-photon Fock states in an experiment relying on a two-detector measurement. In particular, photon bunching ($g^{(2)}(0)>1$) will be observed for low-intensity non-classical light sources in which the presence of the vacuum state is stronger than that of the photon-number Fock state \cite{Faraon2008a}. To resolve the presence of a particular Fock state, it is necessary to evaluate higher-order photon autocorrelation functions and compare them with lower-order ones. For example, a low intensity non-classical light source with a dominant two-photon component will show $g^{(2)}(0)>1$ and $g^{(3)}(0, 0)<1$ (Appendix A). Here, the value of the third-order autocorrelation function \cite{Glauber1963}
\begin{equation}
  g^{(3)}(\tau_1, \tau_2)={\langle a^{\dag} a^{\dag}(\tau_1)a^{\dag}(\tau_1+\tau_2)a(\tau_1+\tau_2)a(\tau_1) a\rangle \over \langle  a^{\dag}a\rangle^3}
\end{equation}
can be estimated with a generalized form of HBT setup that monitors coincidences between three SPCMs and thus allows one to measure the suppression of simultaneous three- and higher-photon events \cite{Kubanek2008}. It is worth noting that the use of photon-number-resolving detectors \cite{Waks2003, Kardynal2008}, especially transition-edge sensors \cite{Lita2008} or superconducting nanowires \cite{Divochiy2008, Steudle2012, Zhou2014}, could provide an alternative technique for the characterization of multi-photon Fock states; however, these devices are still under experimental investigation and are not yet widely available.
 
\begin{figure}
  \centering
  \includegraphics[width=8.5cm]{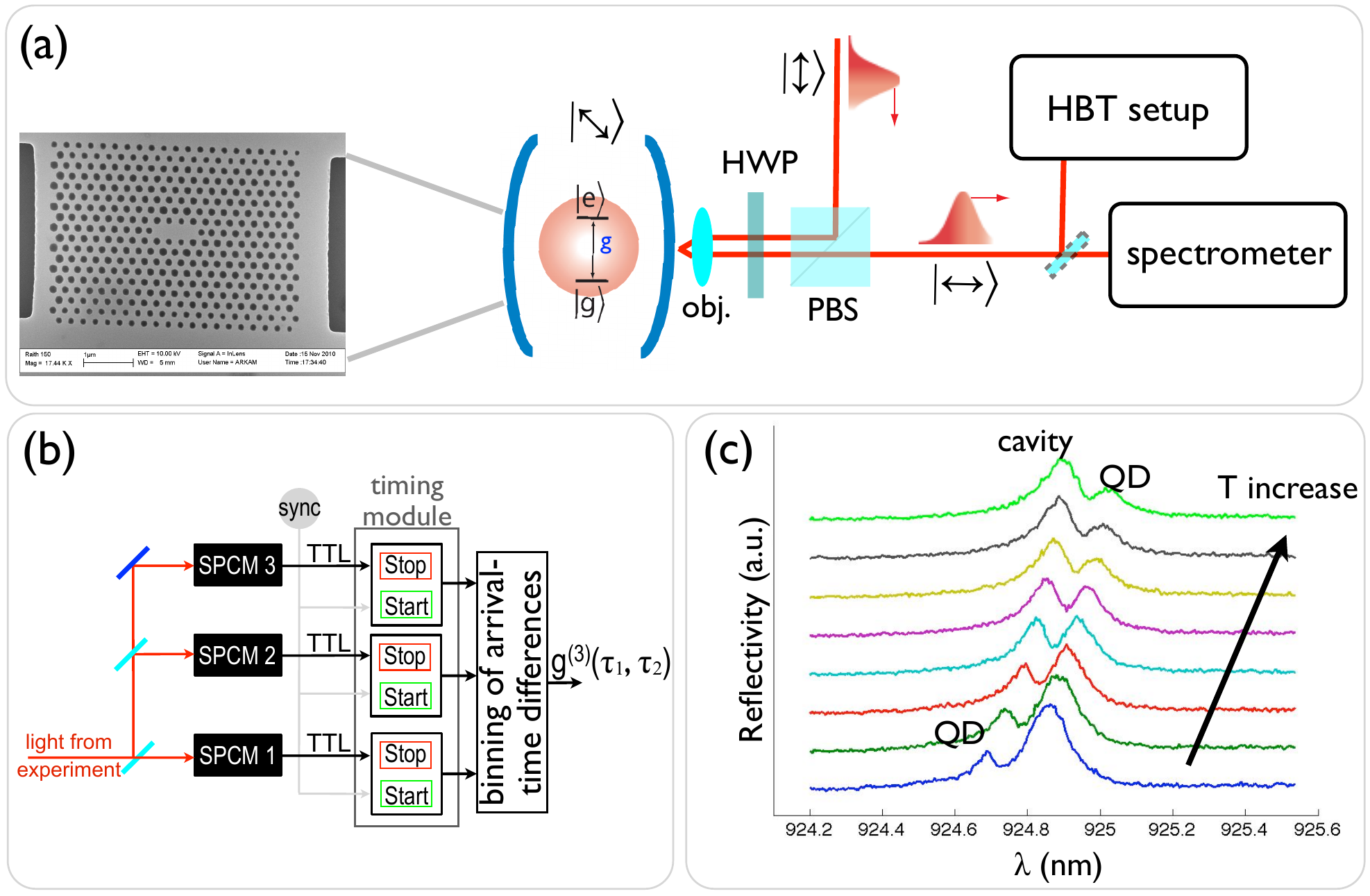}
\caption{(Color online) \textbf{Experimental setup for detecting photon correlations: (a)} A scanning electron microscope image of the photonic-crystal nanocavity (left) and the schematics of the cross-polarized microscopy setup \cite{Faraon2008a}.  The polarizing beam splitter (PBS) in combination with the half-wave plate (HWP) allows us to filter and select only the light that circulated inside the cavity. \textbf{(b)} Schematics of the generalized HBT setup used to detect arrival times of three-photon events, from  which the third-order autocorrelation function $g^{(3)}(\tau_1, \tau_2)$ is then extracted. \textbf{(c)} As the QD is temperature tuned across the resonance of the cavity, an anti-crossing is observed in the system's spectrum (the cross-polarized reflectivity curves are obtained by using a superluminescent broadband diode as the source).}
\label{fig_setup}
\end{figure}


Here, we report the observation of non-classical third- and fourth-order photon correlations in an originally coherent probe after it was transmitted through a semiconductor nanocavity containing a single quantum emitter. Our system consists of a self-assembled InAs quantum dot (QD) embedded in a three-hole linear defect nanocavity (L3) \cite{Akahane2003} in a two-dimensional GaAs photonic crystal, fabricated as described in previous work \cite{Englund2007}. The QD--cavity system is maintained at cryogenic temperatures (between 4K and 50K) using a continuous-flow liquid helium cryostat. We excite this system with focused pulses from a mode-locked Ti:Sapph laser tuned near the bare cavity resonance, and the emitted light was collected with a high numerical aperture objective lens (NA = 0.75). We employ the cross-polarized reflectivity technique (input source orthogonal to the collected reflected signal, and at $45^\circ$ relative to the cavity mode) that mimics the results of a transmission measurement \cite{Faraon2008a}. This experimental setup is depicted in Fig.~\ref{fig_setup}(a).

\begin{figure}[t]
  \centering
  \includegraphics[width=8.5cm]{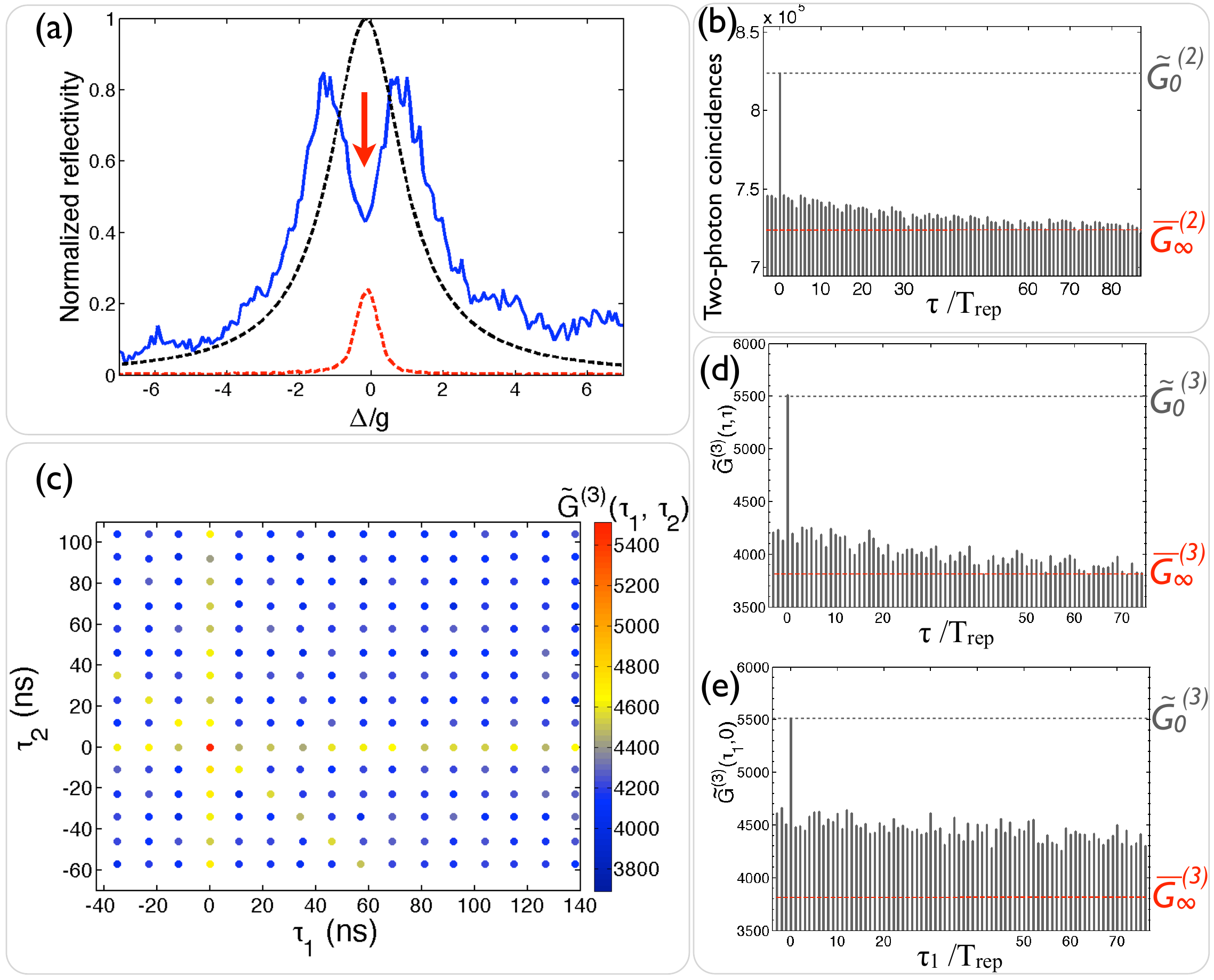}
\caption{(Color online) \textbf{Correlations in a resonantly probed strongly-coupled dot--cavity system: (a)} The transmission spectrum of the strongly coupled system in which the quantum dot was tuned into resonance with the cavity. For comparison purposes, the dashed grey curve plots the calculated transmission of an empty cavity with $Q=6,200$, while the red dotted curve represents the spectrum of the probe pulse (whose center frequency is tuned into resonance with the dot and the cavity, as marked by a red arrow).  \textbf{(b)} Two-photon coincidence counts $\tilde{G}^{(2)}(\tau)$ observed in the transmission of the strongly-coupled system. Notice the classical bunching caused by QD blinking. The second-order autocorrelation is $\bar{g}^{(2)}(0)=\tilde{G}^{(2)}_{0}/ \bar{G}^{(2)}_{\infty}=1.141\pm0.003$. \textbf{(c)} Three-photon coincidence counts $\tilde{G}^{(3)}(\tau_1,\tau_2)$ observed in the photons transmitted through the resonantly probed system. \textbf{(d)} Diagonal elements of the $\tilde{G}^{(3)}(\tau_1, \tau_2)$, with $\tau_1=\tau_2=\tau$ The QD blinking again results in classical bunching that decays with the same time scale as for that observed in $\tilde {G}^{(2)}(\tau)$. \textbf{(e)} $\tilde{G}^{(3)}(\tau_1, \tau_2)$, with $\tau_2=0$, corresponding to the three-photon events in which the system emits the second and third photons simultaneously. }
\label{fig_SC-g3}
\end{figure}

We verify the strong coupling between the cavity and the QD by observing an anti-crossing in the reflectivity (taken using a superluminescent broadband diode as a source) when the QD is temperature tuned through resonance with the cavity (Fig.~\ref{fig_setup}(c)). By fitting the observed spectrum (Fig.~\ref{fig_SC-g3}(a)) (see Appendix A) and assuming that the QD radiative and dephasing rates are $\gamma/2\pi=\gamma_d/2\pi=1$GHz based on previous work \cite{Majumdar2012a}, we extract the experimental parameters of the system -- the decay rate of the cavity field $\kappa/2\pi \approx26$GHz (corresponding to a quality factor $Q\approx6,200$), the QD--cavity coupling rate $g/2\pi\approx21$GHz, and the fraction of the time the dot spends in a dark state (not interacting with the cavity) due to blinking, $p_{dark}\approx0.38$. 
As mentioned earlier, when probed near resonance with coherent light from a laser, such a system can act as an adjustable photon number filter thanks to the anharmonic Jaynes-Cummings ladder \cite{Majumdar2012a}. Prior to this work however, only measurements of the second-order autocorrelation function have been performed on such a system \cite{Faraon2008a, Reinhard2012}. 

The relevant features of the photon correlations in this system occur at a time-scale given by the lifetime of a photon inside the cavity \cite{Faraon2008a}, which is significantly shorter than the time-resolution of the SPCMs. To resolve these features, we sample the correlations  by a train of pulses from the mode-locked Ti:Sapph laser ($\sim80$ MHz repetition rate). The line width of the original $\sim 3$ ps pulses was reduced to $\Delta \lambda_{FWHM} \approx 0.04$ nm (corresponding to a bandwidth of roughly 14 GHz) by passing the pulses through a monochromator. This allows us to resolve the relevant spectral features of the system while retaining the fast sampling. The average optical power in the pulse train was measured to be $\bar{P}_{probe}\approx0.2$ nW in front of the objective lens, which at an $f_{rep}\sim80$ MHz repetition rate and with a coupling efficiency of $\eta\sim 0.01$ corresponds to an approximate intra-cavity photon number (during the on-time of the pulse) of $n = \frac{\eta \bar{P}_{probe}}{f_{rep} \hbar \omega} \approx 0.12$.

We first tune the pulses to be on resonance with the QD--cavity system (red arrow in Fig.~\ref{fig_SC-g3}(a)) and record the arrival times of the transmitted photons using the three-SPCM setup from Fig.~\ref{fig_setup}(b), with the system held at a temperature of $T\approx30$ K. From this data we can extract the second-order autocorrelation function $g^{(2)}(\tau)$ via the two-photon coincidence counts $\tilde{G}^{(2)}(\tau)$ (here $\tilde{G}^{(2)}$ means we have time-binned the raw detection events but not yet normalized them) detected between SPCM1 (start) and SPCM2 (stop) (Fig.~\ref{fig_SC-g3}(b)). In addition to two-photon bunching caused by photon-induced tunneling, this data also reveals the presence of classical bunching resulting from quantum dot blinking \cite{Faraon2008a, Santori2004}. The latter manifests itself as an exponential decay of coincidence counts $\tilde{G}^{(2)}(\tau)$ for increasing $\tau$. The time constant of this decay, $T_{decay}\approx0.5$ $\mu$s, is much longer than the decay of the bunching from photon-induced tunneling, and we extract it together with the normalization constant $\bar{G}^{(2)}_{\infty}$ by fitting the histogram with the function $\tilde{G}^{(2)}(mT_{rep}) = (\bar{G}^{(2)}(T_{rep})-\bar{G}^{(2)}_{\infty}) e^{-mT_{rep}/T_{decay}}+\bar{G}^{(2)}_{\infty}$. Note that this decay time is determined by the mean switching rate between the bright and dark states, and is independent of the fraction of time $p_{dark}$ the quantum dot actually spends in the dark state. A more detailed analysis of the dark state dynamics can be obtained using a rate equation approach \cite{Davanco2014}, but that is beyond the scope of this work. Unfortunately, because of the much faster time-scales of our system compared to conventional atom--cavity experiments \cite{Koch2011}, we cannot resolve the decay rate of photon bunching caused by photon-induced tunneling, as this happens within the time scale of the individual pulses. After normalization, the second-order autocorrelation is $\bar{g}^{(2)}(0)=\tilde{G}^{(2)}_{0}/ \bar{G}^{(2)}_{\infty}=1.141\pm0.003$ (we use the notation $\bar{g}^{(2)}$ to indicate we have both time-binned and normalized the raw coincidence counts, so this represents our experimental measurement of the theoretical value $g^{(2)})$.
 
In a process analogous to obtaining the second-order autocorrelation function, we now extract the third-order temporal auto-correlation function $\bar{g}^{(3)}(\tau_1, \tau_2)$ via the three-photon coincidence counts  $\tilde{G}^{(3)}(\tau_1,\tau_2)$ shown in Fig.~\ref{fig_SC-g3}(c). The observed correlations in this plot are in agreement with the previously reported measurements of $g^{(2)}(\tau)$ in the photon tunneling regime of a strongly coupled QD--cavity system. In particular, the noticeable lines of enhanced peaks correspond to the number of three photon events in which (i) the first and second photon arrive simultaneously (the vertical line, $\tau_1=0$), (ii) the second and third photon arrive simultaneously (the horizontal line, $\tau_2=0$, shown in more detail in Fig.~\ref{fig_SC-g3}(e)), and (iii) the first and third photon arrive simultaneously (the diagonal line with $\tau_1+\tau_2=0$). At the same time, the highest peak at $(\tau_1, \tau_2) = (0, 0)$ corresponds to third-order temporal bunching in transmitted photons with $\bar{g}^{(3)}(0, 0) =1.45 \pm 0.04$, which is noticeably more than the value obtained for $\bar{g}^{(2)}(0)$.  Note that the range of variation in this value has contributions both from the standard deviation of the number of detection events for a given pulse (as determined by a Poissionian distribution) and from the uncertainty in the normalization constant -- this is discussed in more detail in Appendix A.

In contrast, the same $g^{(3)}$ and $g^{(2)}$ measurements for the photoluminescence from a QD weakly coupled to a photonic-crystal nanocavity show a very different signature \cite{Stevens2014}. Namely, in such a system we discover that $\bar{g}^{(3)}(0, 0) < \bar{g}^{(2)}(0)$, i.e., there are significantly fewer events in which three photons arrive simultaneously than events in which two photons arrive simultaneously, as expected for an imperfect single photon source (see Appendix B).
\begin{figure*}[t]
  \centering
  \includegraphics[width=18cm]{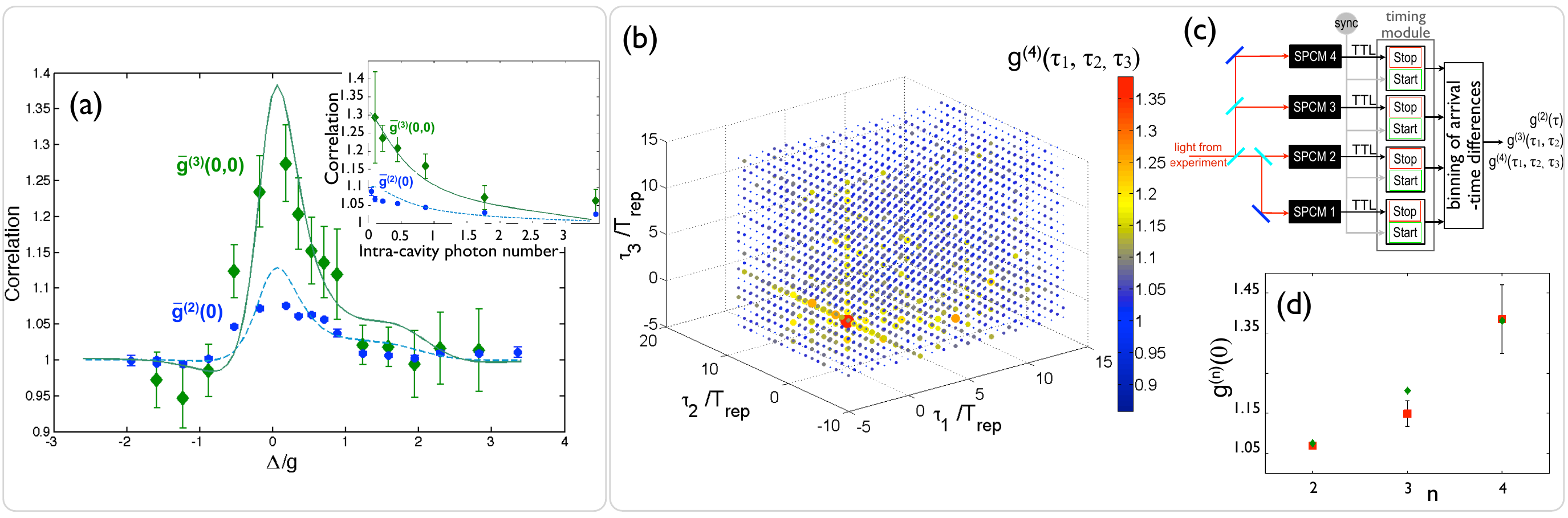}
\caption{(Color online) \textbf{(a)} A frequency scan of the autocorrelation measurements with $\bar{P}_{probe}\approx0.3$ nW, corresponding to intracavity photon number $n \approx 0.17$ (the asymmetry of the data is due to the dot being slightly detuned from the cavity). The inset shows a comparison of the second- and third-order autocorrelations for various levels of $\bar{P}_{probe}$, when the probe is on resonance with the QD and the cavity. The solid (dashed) lines plot the result of a numerical simulation for the second- (third-) order correlations. \textbf{(b)} A visualization of the time-binned and normalized fourth-order autocorrelation function $\bar{g}^{(4)}(\tau_1, \tau_2,\tau_3)$. To guide the eye, the value of each peak is represented both by color and size of the plotted data point.   \textbf{(c)} Schematics of the expanded HBT setup used to detect arrival times of up to four-photon events used to obtain autocorrelation functions up to the fourth-order, $\bar{g}^{(4)}(\tau_1, \tau_2,\tau_3)$ shown in (b). \textbf{(d)} Increasing values of the autocorrelation functions $\bar{g}^{(n)}$ at zero time delay(s), plotted as a function of their order $n$ (the red squares with error bars represent experimental data, while the green diamonds plot the results of a numerical simulation). To obtain a sufficient number of four-photon coincidences over a reasonable  data collection time, the system was probed with $\bar{P}_{probe}\approx1.0$ nW, which partially saturated the dot and resulted in lower observed values of  $\bar{g}^{(3)}(0,0)$ and $\bar{g}^{(2)}(0)$ in this particular measurement. }
\label{fig_freq-scan}
\end{figure*}

We repeat the autocorrelation measurements for a set of probe laser frequencies to map the spectral dependence of $\bar{g}^{(3)}(\tau_1, \tau_2)$. Because the cavity had slightly shifted in frequency, the measurement was now performed with the sample kept at a higher temperature ($\sim40$ K instead of $\sim30$ K), which negatively affected the amount of detectable photon bunching. The QD is also slightly red-detuned from the cavity resonance. Nevertheless, the frequency scan in Fig.~\ref{fig_freq-scan}(a) shows that as the probe is tuned, the third-order autocorrelation $\bar{g}^{(3)}(0, 0)$ of the transmitted photons exhibits either antibunched or bunched behavior as the system transitions from the photon blockade to the photon-induced tunneling regime (for a probe red-detuned from the cavity resonance). For comparison, Fig.~\ref{fig_freq-scan}(a) also shows the values of $\bar{g}^{(2)}(0)$ obtained for the same frequency scan. In the tunneling regime $\bar{g}^{(3)}(0, 0) > \bar{g}^{(2)}(0)$, i.e.\ the simultaneous arrival of three photons is enhanced compared to simultaneous two-photon arrivals, which is in qualitative agreement with numerical simulations (shown in Appendix A). During the experiment, we kept the probe power constant at $\bar{P}_{probe}\approx0.3$ nW (corresponding to an approximate intra-cavity photon number of 0.17) and the coupling of the probe into the cavity was re-optimized for every data point. The data in Fig.~\ref{fig_freq-scan} show a good agreement with numerical simulations of the values of ${g}^{(2)}(0)$ and $g^{(3)}(0, 0)$ as a function of probe detuning, given the system parameters measured earlier ($g/2\pi=21$ GHz, $\kappa/2\pi=26$ GHz, and $\gamma/2\pi=\gamma_d/2\pi=1$ GHz), a probe driving strength of $E/2\pi = 10$ GHz, a QD--cavity detuning of $\Delta=20$ GHz, and the fraction of QD ``dark state'' time $p_{dark}\approx0.9$ (note that this is significantly higher than our earlier estimate based on the reflectivity spectrum of the QD-cavity system, probably due to the higher temperature needed to bring the dot into resonance with the cavity during this measurement). Importantly, the experimental data and numerical simulations show that the bunching in ${g}^{(2)}(0)$ and $g^{(3)}(0, 0)$ when the probe is on resonance with the QD and the cavity drops off as a function of probe power (the inset of Fig.~\ref{fig_freq-scan}(a)), leveling out when the intra-cavity photon number nears 1. This power dependence indicates that the bunching we observe is indeed due to photon-induced tunneling and not classical bunching due to blinking; at higher powers, the QD saturates \cite{Majumdar2010}, which in turn diminishes the polariton dip \cite{Englund2012} and reduces the amount of bunching observed in the light transmitted through the system, an effect which is well reproduced by our numerical simulations. Lastly, we would like to point out that there are several factors that account for the difference between the theoretically predicted values of the second- and third-order autocorrelations $\bar{g}^{(2)}(\tau)$ and $\bar{g}^{(3)}(\tau_1, \tau_2)$ and our experimentally observed values: background light due to imperfect extinction of the uncoupled probe in the cross-polarization setup, quantum dot blinking and spectral diffusion, temperature-dependence of the quantum dot dephasing rate, and non-negligible bandwidth of the probe pulses. In particular, the quantum dot blinking and background light cause the observed signal to have a large coherent-state component.

We observe that the theoretically predicted values of $g^{(3)}(0, 0)$, the third-order photon correlations in a laser beam transmitted through a strongly-coupled QD--cavity system, differ more significantly than the values of $g^{(2)}(0)$ from the unity expected from coherent (laser) light. This confirms that $g^{(3)}(\tau_1, \tau_2)$ is a more sensitive diagnostic tool for observing non-classicality in the measured photon statistics, as it can more clearly be distinguished from the signature of coherent light. This approach -- increasing the order of the correlations in order to get a more clear signature of non-coherent light -- is further illustrated in Fig.~\ref{fig_freq-scan}(b-d) showing the preliminary results of our measurements of the fourth-order autocorrelation function
\begin{widetext}
\begin{equation}
  g^{(4)}(\tau_1, \tau_2,\tau_3)=\\{\langle a^{\dag} a^{\dag}(\tau_1)a^{\dag}(\tau_1+\tau_2)a^{\dag}(\tau_1+\tau_2+\tau_3)a(\tau_1+\tau_2+\tau_3)a(\tau_1+\tau_2)a(\tau_1) a\rangle \over \langle  a^{\dag}a\rangle^4}
\end{equation}
\end{widetext} 

The four-photon correlations (Fig.~\ref{fig_freq-scan}(b)) were obtained by adding another photon counter to the generalized HBT setup (Fig.~\ref{fig_freq-scan}(c)), and then binning and normalizing the four-photon coincidences in the transmitted light through this QD-cavity system probed with resonant (Fig.~\ref{fig_SC-g3}(a)) laser pulses, a process equivalent to that described above for $g^{(2)}(\tau)$ and $g^{(3)}(\tau_1,\tau_2)$. Fig.~\ref{fig_freq-scan}(d) then shows how value of the autocorrelation-function at zero time keeps increasing with the order of the auto-correlation function for light transmitted by the cavity-QD in photon-induced tunneling regime. A down-side of this approach is the increasing measurement time required to collect enough events for a meaningful statistical analysis, which might not be possible in some experimental systems (see Appendix A for a discussion of the count rate in our setup).

Finally, to achieve efficient generation of photon pairs and other higher-order Fock states in this way, a system with a better dot--cavity coupling strength $g$ and higher cavity quality factor would be needed (as indicated by the numerical simulations presented in Appendix A). In addition, optimizing the dot--cavity detuning of the system with current parameters could possibly also be employed to improve photon blockade and as an alternative scheme for generating higher-order Fock states, as recently proposed by S\'{a}nchez-Mu\~{n}oz \textit{et al.}\ \cite{Munoz2013}. A source of such higher-order photon states could then be used for efficient generation of the highly-entangled NOON-states, which are particularly interesting for quantum metrology and high resolution quantum lithography and sensing \cite{Afek2010}.
Lastly, these higher-order autocorrelations have the potential to be used for monitoring phase transitions in condensed matter simulations based on photon gases  \cite{Carusotto2012}.

This work was supported by the Army Research Office, by the Air Force Office of Scientific Research, MURI center for multi-functional light-matter interfaces based on atoms and solids, and by DARPA. Additional support was provided by the Stanford Graduate Fellowship (K.F.), the National Science Foundation Graduate Research Fellowship (S.B.), and the Swiss National Science Foundation (K.G.L.). Work was performed in part at the Stanford Nanofabrication Facility (SNF) of NNIN supported by the National Science Foundation, and at the Stanford Nano Center (SNC).

\appendix

\section{Methods}

\begin{figure*}
  \centering
  \includegraphics[width=14cm]{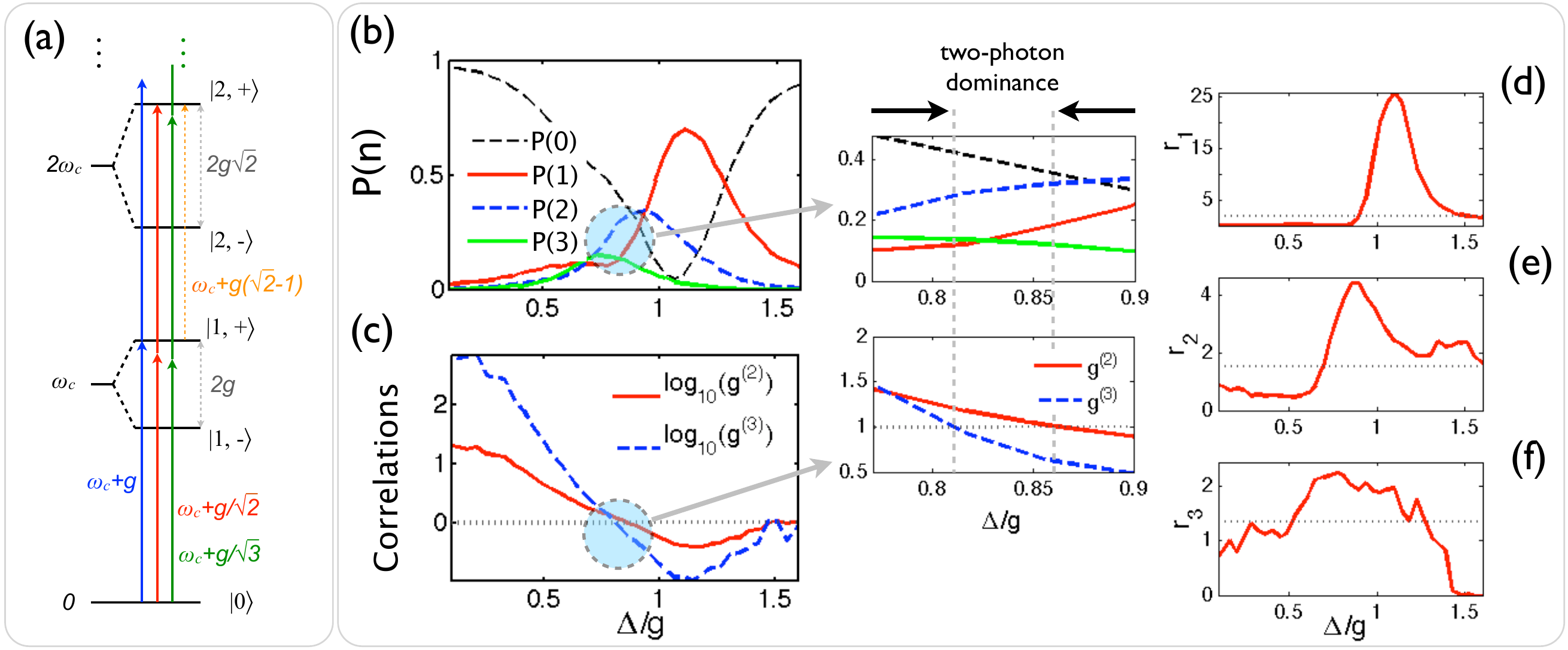}
\caption{(Color online) \textbf{Creating and detecting two-photon states inside a cavity containing a quantum emitter: (a)} Energy diagram of a strongly-coupled QD--cavity system showing the spacings between the levels of the first- and second-order manifolds. Because the energy differences between consecutive levels are not constant, the system can act as a photon number filter when the frequency of the probe laser is tuned properly. Here, the blue arrow represents the probe frequency at which only individual photons couple into the system. The red arrow represents the probe frequency at which photons couple in pairs via a two-photon transition, while the green arrow represents the frequency at which a three-photon transition is addressed. \textbf{(b)} Probability of  $n$ photon state, $P(n)$, inside the QD-cavity system as a function of laser-cavity detuning $\Delta$ and \textbf{(c)} the corresponding second- and third-order correlation functions (plotted in log scale).  The photon statistics was numerically calculated for a system driven by Gaussian pulses with duration  $\tau_p \sim 25$ ps. The simulation parameters for both (b) and (c) are $g = 2\pi \times 40$ GHz, $\kappa =2\pi \times 4$ GHz,  and $\mathcal{E}_o=2\pi \times 9$GHz  which are close to the highest achievable $g$ with this type of cavity and quantum dot and to the highest quality factor ($Q\approx 25,000$) measured in our laboratory; $\gamma/2\pi=1$ GHz and pure QD dephasing $\gamma_d$ is neglected. The zoomed in sections of the plots (both in linear scale) show the frequency range (marked by the vertical dashed lines) in which the two-photon state is dominant over the other states ($P(2)>P(1)+P(3)$) and $g^{(2)}(0)>1$ and $g^{(3)}(0, 0)<1$.
The functions  \textbf{(d)} $r_1= {P^2(1)\over P(0)P(2)}$,  \textbf{(e)}  $r_2= {P^2(2)\over P(1)P(3)}$, and  \textbf{(f)}  $r_3 ={P^2(3)\over P(2)P(4)}$, 
as a function of frequency of the probe laser. 
Each plot is contrasted with  $r_1$, $r_2$ and $r_3$ for a coherent light source with $\langle a^{\dag}a\rangle=2$ (dashed gray lines).
}
\label{fig_sims}
\end{figure*}

\subsection{Theoretical Modeling:}
Our simulations were performed by numerically integrating the quantum master equation using the QOToolbox originally developed by Tan \cite{qotoolbox}. We model the  dynamics of a coupled QD--cavity system (coherently driven by a laser field) with the Jaynes-Cummings Hamiltonian of the form:

\begin{multline}
  \label{eqn:H} H=\Delta_a \sigma_+ \sigma_-+\Delta_ca^\dag a \\
  +ig(a^\dag\sigma_--a\sigma_+)+\mathcal{E}(t)a+\mathcal{E}^*(t)a^\dag,
\end{multline}

\noindent which assumes the rotating wave approximation (RWA) and a frame of reference rotating with the frequency of the laser field $\omega_l$. Here $\Delta_a=\omega_a-\omega_l$ and $\Delta_c=\omega_c-\omega_l$ are respectively the detuning of the QD resonant frequency $\omega_a$ and the cavity resonance frequency $\omega_c$ from the laser, $g$ is the coherent coupling strength between the QD and the cavity mode, $\mathcal{E}(t)=\sqrt{\kappa P(t)\over{\hbar \omega_c}}$ is the slowly varying envelope of the coherent driving field with power $P(t)$ incident onto the cavity, and $a$ is the annihilation operator for the cavity mode. If the excited and ground states of the QD are denoted by $|e\rangle$ and $|g\rangle$ then $\sigma_-=|g\rangle\langle e|$ and $\sigma_+=|e\rangle\langle g|$.

Three main loss mechanisms of this system (the cavity field decay rate $\kappa=\omega_{c}/2Q$ where $Q$ is the quality factor of the resonator, QD spontaneous emission rate $\gamma$, and pure dephasing of the QD $\gamma_d$) are incorporated in the master equation:
\begin{equation}
  \label{Maseq} \frac{d\rho}{dt}=-i[H,\rho]+ \kappa
  \mathcal{L}[a]+\gamma \mathcal{L}[\sigma]+\gamma_d \mathcal{L}[\sigma_+\sigma_-],
\end{equation}

\noindent where $\rho$ is the density matrix of the coupled QD--cavity system and $\mathcal{L}[D]$ is the Lindblad operator corresponding to
operator $D$, defined as

\begin{equation}
  \mathcal{L}[D]= 2D\rho D^\dag-D^\dag D \rho-\rho D^\dag D.
\end{equation}

We define the values of the second- and third-order autocorrelation functions $g^{(2)}(\tau_1)$ and $g^{(3)}(\tau_1, \tau_2)$ \cite{Glauber1963} as
\begin{align}
  &g^{(2)}(\tau_1)={\langle a^{\dag} a^{\dag}(\tau_1) a(\tau_1) a\rangle \over \langle  a^{\dag}a\rangle^2} \nonumber \\
  &g^{(3)}(\tau_1, \tau_2)={\langle a^{\dag} a^{\dag}(\tau_1)a^{\dag}(\tau_1+\tau_2)a(\tau_1+\tau_2)a(\tau_1) a\rangle \over \langle  a^{\dag}a\rangle^3},
\end{align}
where $\tau_1$ is the time between the arrival of the first and second photons and $\tau_2$ is the time between the arrival of the second and third photons. Thus, if we assume that the output photon state transmitted through the system can be expressed as a superposition of the Fock (photon number) states $\left| \psi \right\rangle = \sum_n c_n \left| n \right\rangle$ where the probability of the $n^{\mathrm{th}}$ Fock state is $P(n) = \left| c_n \right|^2$, it follows that
\begin{align}
  &g^{(2)}(0) =  \frac{\sum_n n(n-1)P(n)}{[\sum_n nP(n)]^2} \nonumber \\
  &g^{(3)}(0,0) = \frac{\sum_n n(n-1)(n-2)P(n)}{[\sum_n nP(n)]^3}
\end{align}
In particular, this means that $g^{(2)}(0)=0$ for a single-photon pulse train (whether perfect or sparse), while $g^{(2)}(0)=1/2$ for a perfect two-photon pulse train but $g^{(2)}(0)\approx{1/2 \varepsilon^2}$ for a sparse two-photon pulse train (defined as $\left| \psi \right\rangle \approx \sqrt{1-\varepsilon^2}\vert 0 \rangle + \varepsilon \vert 2 \rangle$).

Note that a classical light source producing a coherent state $\vert \alpha\rangle = \sum_n {\alpha^n \over \sqrt{n!}} \vert n\rangle$ can also have a particular Fock state $\vert m \rangle$ to be the state with the highest probability of occurrence $P(m)$, if $\alpha$ is chosen such that $m+1>\alpha^2>m$. Thus to evaluate the non-classicality of a Fock-state generating light source, we define the ratio $r_m ={P^2(m)\over P(m-1)P(m+1)}$ \cite{Majumdar2012a}, which contrasts the probability of state $\vert m \rangle$ with the probabilities of the neighboring Fock states $\vert m-1 \rangle$ and $\vert m+1 \rangle$ in a given light source.  
For a coherent state the ratio $r_m =1+1/m$ remains a constant for a given $m$, which cannot be optimized by adjusting the value of $\alpha$. This in turn leads to $g^{(n)}(0)=1$ for all $n$ for an $m$-photon Fock-state generating light source based on attenuation of coherent light (based on the simple application of the expressions above for $g^{(2)}$ and $g^{(3)}$). Thus the level of non-classicality of a weak $m$-photon Fock-state generating light source, such as one based on the generalized photon blockade, can be quantified either by how much its $r_m$ differs from the classical limit (Fig.~\ref{fig_sims}(d-f)) or by looking at the values of its $g^{(m)}(0)$ and $g^{(m+1)}(0)$ (Fig.~\ref{fig_sims}(c)).

Fig.~\ref{fig_sims}(a) shows the theoretical technique for addressing higher-order manifolds of the Jaynes-Cummings ladder of a strongly coupled quantum dot--photonic crystal nanocavity system \cite{Faraon2008a}, while Fig.~\ref{fig_sims} (b-f) plot the numerical simulation results that show the response of the system to different probe laser frequencies. The parameters used for the simulation are those of an ideal QD-cavity system that is currently within experimental reach and that is driven with laser pulses comparable to those used in our experiment. 
As expected, the system behaves as an highly non-classical source in the single-photon regime (blue arrows in Fig.~\ref{fig_sims}(a)). The single-photon component dominates the vacuum, as well as all multi-photon Fock states (Fig.~\ref{fig_sims}(b)), resulting in $g^{(2)}(0)\approx0.4$ (Fig.~\ref{fig_sims}(c)) and $r_2\approx25$ (Fig.~\ref{fig_sims}(d)) for $\Delta/g\approx1.1$.   
Accessing the two-photon regime requires addressing the levels of the second manifold of the Jaynes-Cummings ladder (red arrows in Fig.~\ref{fig_sims}(a)). This addressing is more difficult to do precisely, given that the linewidth of the levels in the second manifold is roughly twice as big as the linewidth of the levels in the first manifold\cite{Bajcsy2013}. As a result, the frequency for the maximum probability of a two-photon state ($\Delta/g=0.94$, Fig.~\ref{fig_sims}(b)) does not fully coincide with the maximum nonclassicality of the two-photon regime ($r_2\approx 4.4$ at $\Delta/g\approx0.9$, Fig.~\ref{fig_sims}(e)),  and both are actually outside of the frequency region in which the two-photon state dominates over the other non-zero Fock states ($g^{(3)}(0,0)<1$ and $g^{(2)}(0)>1$, zoomed section of Fig.~\ref{fig_sims}(c)).
The detrimental effect of the increasing level broadening of the higher-order manifolds fully takes over when one tries to access the three-photon regime (green arrows in Fig.~\ref{fig_sims}(a)). While $P(3)$ is maximized at  $\Delta/g\approx0.7$ (Fig.~\ref{fig_sims}(b)), the three-photon state is far from dominant and $r_3$ only reaches $\sim2$ (Fig.~\ref{fig_sims}(f)).

These results illustrate the main limit of this scheme for non-classical light generation, which comes from the unresolvability of the higher-order manifolds in the currently achievable GaAs L3 cavities with self-assembled QDs. The full potential of this scheme, in particular its photon-number filtering capabilities, could however be utilized in optical systems with higher quality factors (such as silicon-based L3 cavities), smaller mode volumes (which can potentially result in higher coupling strengths), or in circuit cQED systems.     

\subsection{Extraction of parameters from the spectrum of a strongly-coupled system:}
Based on our previous work \cite{Majumdar2012a}, we estimate QD dipole decay and pure dephasing rates to be $\gamma/2\pi \approx \gamma_d/2\pi \approx1$ GHz (for simplicity we neglect any temperature-dependence of the dephasing). To describe the QD's blinking and spectral diffusion, we use a simplified model in which we assume the QD is in a bright state and resonant with the cavity for $p_{bright}$ fraction of the time and does not interact with the cavity for $p_{dark}=1-p_{bright}$ fraction of the time, when it either goes dark \cite{Akimov2007} or has jumped to a far off-resonant state. To extract the decay rate of the cavity field $\kappa$ and the QD--cavity coupling rate $g$, we perform a least-squares fit of $F_{total}=p_{bright}I_{in}F_{DIT}+(1-p_{bright})I_{in}F_{cav}+I_{bg}$ to the observed transmission spectrum (Fig. 3(a)) of the strongly coupled system. Here, $I_{in}$ is the intensity of the laser coupled into the cavity, $I_{bg}$ is the background laser signal unsuppressed by the cross-polarization setup, 
\begin{equation}
  F_{DIT}=\left \vert \frac{\kappa [\gamma_{tot}+i(\omega_{QD}-\omega_l)]}{[\kappa+i(\omega_{cav}-\omega_l)][\gamma_{tot}+i(\omega_{cav}-\omega_l)]+g^2} \right \vert^2
\end{equation}
is the transmission spectrum of a strongly-coupled QD--cavity system in the weak probe approximation \cite{Englund2007},
\begin{equation} 
  F_{cav}=\left \vert \frac{\kappa}{\kappa+i(\omega_{cav}-\omega_l)} \right \vert^2
\end{equation}
is the Lorentzian transmission spectrum of an empty cavity, $\gamma_{tot}=\gamma+\gamma_d$, $\omega_{cav}$ and $\omega_{QD}$ are (respectively) the resonant frequencies of the cavity and the dot, $\omega_l$ is the frequency of the probe laser, and we used $p_{bright}$, $I_{in}$, $I_{bg}$, $\kappa$, $g$, $\omega_{cav}$, and $\omega_{QD}$ as the fitting parameters.

\subsection{Calculation of $g^{(3)}(\tau_1, \tau_2)$:}
For the measurement of $g^{(3)}(\tau_1, \tau_2)$, we implemented a three-channel photon arrival detecting setup (shown in Fig.~2(b)) that records the arrival times of individual photons at each of the three SPCMs using FPGA-based timing electronics. We time-bin the resulting data files containing the arrival sequence in order to produce a time-of-arrival histogram for three-photon events. Similar to the process used to extract the number of two-photon coincidences $G^{(2)}(\tau)$ from a two-detector measurement, our time-binning algorithm uses a photon detected by SPCM1 as a start signal, the detection of a photon by SPCM2 as the first stop ($\tau_1$), and the detection of a photon by SPCM3 as the second stop ($\tau_1+\tau_2$). This results in a 2-D histogram for $G^{(3)}(\tau_1, \tau_2)$ with a grid of peaks with spacing given by the repetition rate of the Ti:Sapph pulses, as shown for instance in Fig.~\ref{fig_WC-g3}(a). The width of the peaks in the grid is in this case mostly given by the time jitter of the TTL output from the SPCMs, so integrating each peak will result in a time-binned average number of three-photon events, which we denote by $\tilde{G}^{(3)}(\tau_1, \tau_2)$ (examples plotted in Fig.~\ref{fig_WC-g3}(b-d)).

To obtain the normalized pulse-averaged third-order autocorrelation function $\bar{g}^{(3)}(\tau_1, \tau_2)$, we generalize the procedure for extracting the value of the second-order autocorrelation function \cite{Faraon2008a} -- specifically, we rescale the data such that $\bar{g}^{(3)}(\tau_1\rightarrow \infty, \tau_2\rightarrow \infty)=1$. This is done by dividing $\tilde{G}^{(3)}(\tau_1, \tau_2)$ by $\bar{G}^{(3)}_{\infty}$, which we 
obtain by fitting the histogram of $\tilde{G}^{(3)}(\tau, \tau)$ with the function
\begin{multline}
  \tilde{G}^{(3)}(mT_{rep}, mT_{rep}) = \\
  \left(\bar{G}^{(3)}(T_{rep}, T_{rep})-\bar{G}^{(3)}_{\infty}\right) e^{-mT_{rep}/T_{decay}}+\bar{G}^{(3)}_{\infty}
\end{multline}
to remove the effects of probe-induced blinking \cite{Santori2004} on $\bar{G}^{(3)}_{\infty}$.

The error ranges given on our final values for $\bar{g}^{(2)}(0)$ and $\bar{g}^{(3)}(0, 0)$ takes into account both the standard deviation $\sigma_0 = \sqrt{\tilde{G}^{(n)}_0}$ of the number of detection events in the given pulse (at zero time-delay) as derived from Poissonian statistics, as well as the uncertainty $\sigma_\infty$ in the normalization value $\bar{G}^{(n)}_{\infty}$. Combining these effects,
\begin{multline} 
  \bar{g}^{(n)}(\tau_1=0,\tau_2=0,...,\tau_{n-1}=0) \equiv \frac{\tilde{G}^{(n)}_0 \pm \sqrt{\tilde{G}^{(n)}_0}}{\bar{G}^{(n)}_{\infty} \pm \sigma_\infty} \\
  = \frac{\tilde{G}^{(n)}_0 \pm \sqrt{\tilde{G}^{(n)}_0}}{\bar{G}^{(n)}_{\infty}} \left( 1 \mp \frac{\sigma_\infty}{\bar{G}^{(n)}_{\infty}} + \mathcal{O}\left(\frac{\sigma_\infty}{\bar{G}^{(n)}_{\infty}}\right)^2 \right) \\
  \approx \frac{\tilde{G}^{(n)}_0}{\bar{G}^{(n)}_{\infty}} \pm \frac{\left( \bar{G}^{(n)}_{\infty} + \sigma_\infty \right) \sqrt{\tilde{G}^{(n)}_0} + \sigma_\infty \tilde{G}^{(n)}_0}{\left.\bar{G}^{(n)}_{\infty}\right.^2}
\end{multline}
which provides a method of easily computing both the nominal values for $\bar{g}^{(2)}(0)$ and $\bar{g}^{(3)}(0, 0)$ as well as their expected variation.

\subsection{Count rate for multiple-photon correlations}
In our setup, the single-photon count rate on each SPCM was roughly $7\times10^5$ counts per second for the $g^{(4)}$ measurements shown in Fig.~\ref{fig_freq-scan}. After identifying the multiple-photon correlations and time-binning the delays, the coincidence counts are distributed into a histogram consisting of a series of discrete peaks (corresponding to different time delays between photon arrivals in increments of 12.5 ns, the repetition rate of the driving laser), as discussed earlier. On average (away from the zero time-delay peak), each two-photon coincidence peak accumulated counts at a rate of 22,000 to 42,000 counts per second (depending on the particular SPCM configuration), while three-photon coincidence peaks accumulated at a rate of 27 to 42 counts per second, and four-photon coincidence peaks accumulated at a rate of only 0.25 counts per second. It can be seen that moving to the next higher-order autocorrelation function decreases the count rate (and hence increases the integration time) by roughly a factor of 100. Since at minimum several hundred counts are needed to make a reliable measurement, this puts the total integration time for $g^{(4)}$ measurements (in our setup) on the order of hours.

\section{Three-photon correlations in a weakly-coupled QD--cavity system}

To demonstrate the measurement of the third-order autocorrelation function from a solid-state system, we first measured $g^{(3)}(\tau_1, \tau_2)$ from a single photon source based on the spontaneous emission from an individual QD coupled to low quality factor cavity ($Q\sim 2000$). This particular QD--cavity system was in the weakly coupled regime,  with the cavity improving the photon collection efficiency. The QD was temperature-tuned to be on resonance with the cavity at $921$ nm. The system was then illuminated with focused pulses from a mode-locked Ti:Sapph laser tuned to a higher order-mode of the L3 cavity ($\sim889$ nm), which allowed us to excite the dot through a higher-order state \cite{Toishi2009}. The pulses were $\sim3$ ps long with a repetition rate of $\sim80$ MHz. The light emitted by the QD was collected with a high numerical aperture objective (NA = 0.75) and passed through a $\sim1$ nm FWHM band-pass filter to reject the scattered light from the excitation pulses and to suppress any undesired fluorescence.

\begin{figure}
  \centering
  \includegraphics[width=8.5cm]{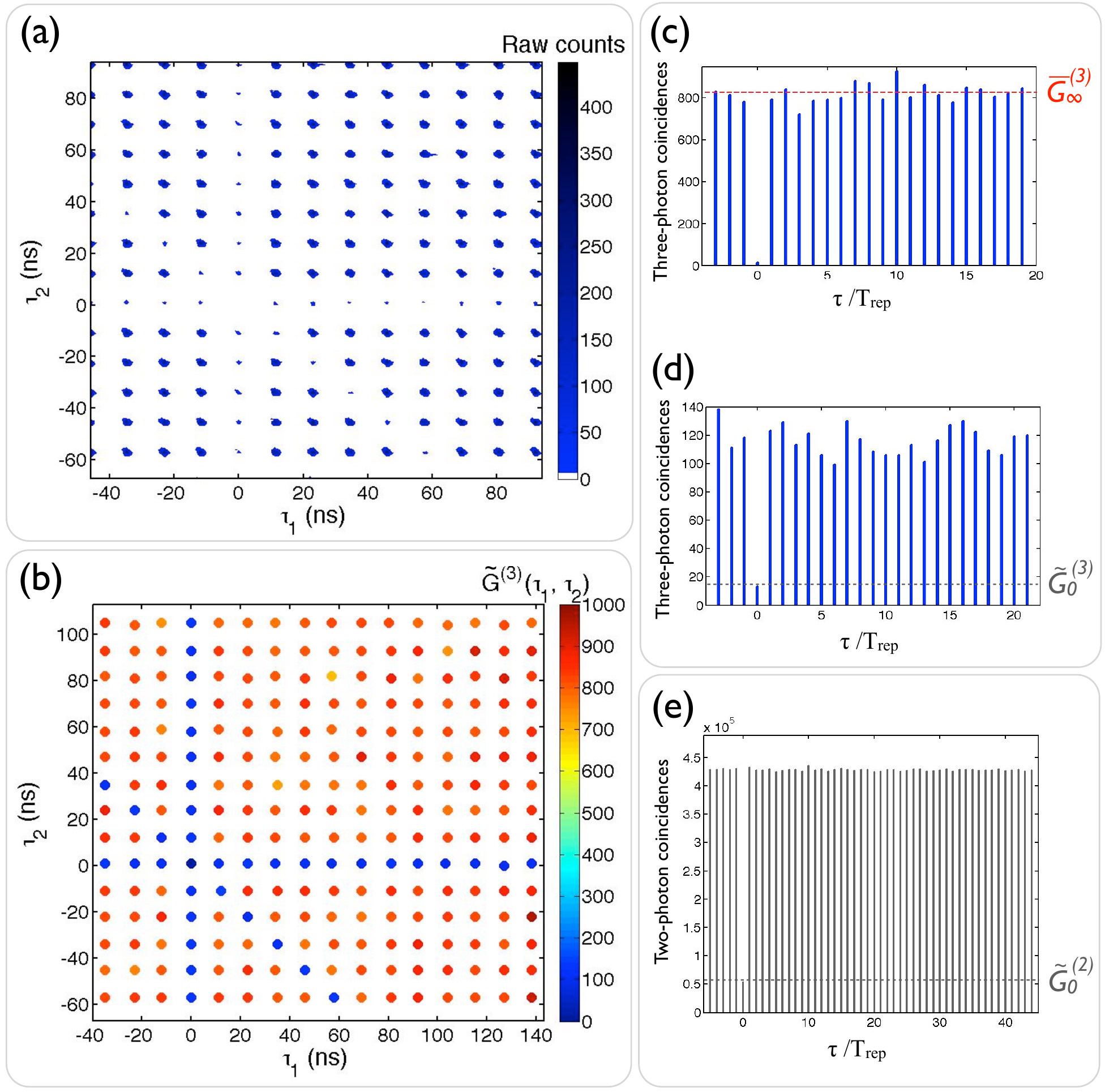}
\caption{(Color online) \textbf{Three photon correlations from a QD fluorescence: (a)} Three-photon coincidences (a subset of the raw data) detected in the fluorescence from a quasi-resonantly ($\sim889$ nm) excited quantum dot weakly coupled to a photonic crystal cavity. Here, $\tau_1$ is the time interval between the arrival of the first and second photon, while $\tau_2$ is the time interval between the arrival of the second and third photon. \textbf{(b)} $\tilde{G}^{(3)}(\tau_1, \tau_2)$, the unnormalized values of the three-photon correlations, obtained by integrating the counts under the coincidence peaks in (a). \textbf{(c)} Diagonal elements of the $\tilde{G}^{(3)}(\tau_1, \tau_2)$, with $\tau_1=\tau_2=\tau$. $T_{rep} \approx12.5$ ns is the repetition period of the pulse train from the Ti:Sapph laser used to excite the quantum dot and the dashed red line marks the normalization level, $\bar{G}^{(3)}_{\infty}=819\pm34$, for the autocorrelation function. \textbf{(d)} Coincidence counts  $\tilde{G}^{(3)}(\tau_1, \tau_2)$, with $\tau_2=0$, corresponding to the three-photon events in which the system emits the second and third photons simultaneously. Note the different scale of the vertical axes compared to (c). \textbf{(e)} Two-photon coincidence counts $\tilde{G}^{(2)}(\tau)$ observed from the QD. The normalization level here is $\bar{G}^{(2)}_{\infty}=(4.26\pm0.01)\times10^5$. }
\label{fig_WC-g3}
\end{figure}

Fig.~\ref{fig_WC-g3}(a) shows a subset of the raw data for the three-photon coincidence histogram collected from the QD fluorescence with our three-channel setup (as discussed in more detail in the Methods section above). Even before any further time-binning, we can see qualitative features of the system that can be intuitively expected for the emission from a single quantum emitter. In particular, the noticeable lines of suppressed peaks correspond to the number of three photon events in which (i) the first and second photon arrive simultaneously (the vertical line, $\tau_1=0$), (ii) the second and third photon arrive simultaneously (the horizontal line, $\tau_2=0$, shown in more detail in Fig.~\ref{fig_WC-g3}(d)), and (iii) the first and third photon arrive simultaneously (the diagonal line with $\tau_1+\tau_2=0$). At the same time, the missing peak at $(\tau_1, \tau_2) = (0, 0)$ corresponds to the number of events in which all three photons arrive simultaneously.

Each peak in the histogram in Fig.~\ref{fig_WC-g3}(a) represents the unnormalized value of the third-order autocorrelation, spread over the duration of the excitation pulse. Since the width of the peaks is an artifact of the SPCM timing jitter, we sum the events under each peak into a single time-bin to obtain the average number of three-photon events $\tilde{G}^{(3)}(\tau_1, \tau_2)$, in which the  photons are spaced by $\tau_1=mT_{rep}$ and $\tau_2=nT_{rep}$  (Fig.~\ref{fig_WC-g3}(b-d)). We find that the normalized third-order autocorrelation at $(\tau_1, \tau_2) = (0, 0)$ is given by $\bar{g}^{(3)}(0, 0) = \tilde{G}^{(3)}(0, 0)/\bar{G}^{(3)}_{\infty}=0.016 \pm 0.005$, i.e., the simultaneous arrival of three photons is almost completely suppressed. For comparison, Fig.~\ref{fig_WC-g3}(e) then plots  the two-photon coincidence counts $\tilde{G}^{(2)}(\tau)$ detected between SPCM1 (start) and SPCM2 (stop), from which we extract  $\bar{g}^{(2)}(0)=\tilde{G}^{(2)}(0)/\bar{G}^{(2)}_{\infty}=0.126 \pm 0.001$. Note that for this case of quasi-resonant excitation we observe the blinking-related decay of two-photon coincidences with $T_{decay}\approx 1.37\mu$s.  The non-zero values of both  $\bar{g}^{(3)}(0, 0)$ and $\bar{g}^{(2)}(0)$ are the result of imperfect spectral filtering of the background photoluminescence (PL) from the sample. We also excited the system through the wetting layer of the quantum dots (860 nm), which resulted in additional PL noise, worsening the single-photon behavior of the system. In this case $\bar{g}^{(2)}(0)=0.795 \pm 0.002$ (with $T_{decay}\approx 0.3\mu$s), while $\bar{g}^{(3)}(0, 0) = 0.59 \pm 0.02$, i.e.\ the number of events in which three photons arrive simultaneously is still significantly lower than the number of events in which two photons arrive simultaneously.

It is also worth mentioning that since we extract the correlations from a complete list of photon arrival times instead of the more conventional approach of detecting two photons and binning the difference of their arrival times, our recorded correlations at $\tau \gg T_{rep}$ are not affected by the exponential decay that otherwise arises as a consequence of the single-stop binning technique \cite{Zwiller2004}.

%

\end{document}